\documentclass[sigconf]{acmart}

\usepackage{booktabs} 
\usepackage{subcaption}
\usepackage{cleveref}

\copyrightyear{2017}
\acmYear{2017}
\setcopyright{acmcopyright}
\acmConference{WiNTECH'17}{October 20, 2017}{Snowbird, UT, USA}\acmPrice{15.00}\acmDOI{10.1145/3131473.3131482}
\acmISBN{978-1-4503-5147-8/17/10}

\begin{document}
\title{Finding by Counting: A Probabilistic Packet Count Model for Indoor Localization in BLE Environments}

\author{Subham De}
\affiliation{%
  \institution{University of Illinois at Urbana-Champaign}
  \city{Urbana}
  \state{Illinois}
  \postcode{61801}
}
\email{de5@illinois.edu}

\author{Shreyans Chowdhary}
\affiliation{%
  \institution{University of Illinois at Urbana-Champaign}
  \city{Urbana}
  \state{Illinois}
  \postcode{61801}
}
\email{shreyanschowdhary2@gmail.com}

\author{Aniket Shirke}
\affiliation{%
  \institution{IIT Bombay}
  \city{Mumbai}
  \country{India}}
\email{aniket8897@gmail.com}

\author{Yat Long Lo}
\affiliation{%
  \institution{University of Illinois at Urbana-Champaign}
  \city{Urbana}
  \country{Illinois}}
\email{yllo2@illinois.edu}

\author{Robin Kravets}
\affiliation{%
  \institution{University of Illinois at Urbana-Champaign}
  \city{Urbana}
  \state{Illinois}
  \postcode{61801}
}
\email{rhk@illinois.edu}

\author{Hari Sundaram}
\affiliation{%
  \institution{University of Illinois at Urbana-Champaign}
  \city{Urbana}
  \state{Illinois}
  \postcode{61801}
}
\email{hs1@illinois.edu}


\renewcommand{\shortauthors}{De et al.}

\begin{abstract}

We propose a probabilistic packet reception model for Bluetooth Low Energy (BLE) packets in indoor spaces and we validate the model by using it for indoor localization. We expect indoor localization to play an important role in indoor public spaces in the future. We model the probability of reception of a packet as a generalized quadratic function of distance, beacon power and advertising frequency. Then, we use a Bayesian formulation to determine the coefficients of the packet loss model using empirical observations from our testbed. We develop a new sequential Monte-Carlo algorithm that uses our packet count model. The algorithm is general enough to accommodate different spatial configurations. We have good indoor localization experiments: our approach has an average error of $\sim 1.2m$, $53\%$ lower than the baseline range-free Monte-Carlo localization algorithm.

\end{abstract}

%
%



\keywords{Internet of Things, Indoor Localization, Bluetooth Low Energy, Probabilistic packet reception model}

\maketitle

\section{introduction}
\label{sec:introduction}

In this paper we develop a probabilistic model for Bluetooth Low Energy (BLE) packet reception within an indoor environment. Then, as a test application, we use the packet reception model for indoor localization. We expect Indoor localization using BLE to play an important role in future retail experiences, facilitating automated checkouts and targeted advertisements.
The importance of developing a packet reception model is two-fold: novel indoor localization techniques; network simulations. First, techniques for indoor localization fall under two camps: Received Signal Strength (i.e. energy loss models) fingerprinting, and range free models that avoid using RSS indicators. We know that RSS indicators to be unreliable---they vary with human presence, presence of obstructions and affected by multi-path loss. Range free models in contrast assume that a heard beacon is within a known distance threshold. A packet reception model allows us to localize based on packet counts without making assumptions on RSS or distance thresholds. Second, a packet reception model would serve as an alternative to RSS based packet models used in network simulators such as NS3.

Our main contributions are a new BLE packet reception model for indoor environments and a sequential Monte-Carlo localization application of the proposed model.  We model the probability of reception of a packet as a generalized quadratic function of distance, beacon power and advertising frequency. We obtain extensive empirical data by conducting experiments varying beacon power and frequency in an experimental testbed with stacks to dampen packet reception. Then, we proposed a Bayesian formulation to determine the coefficients of the packet loss model using the empirical observations. We develop a new sequential Monte-Carlo algorithm that uses our packet count model. The algorithm is general enough to accommodate different spatial configurations.

Our experiments on indoor localization reveal that our proposed approach works well: it has an average error of $\sim 1.2m$ which is $53\%$ lower than the baseline Monte-Carlo Localization algorithm.  Our localization errors within an aisle are even better at $\sim 0.4m$, with the increased errors arising due to the transition.

In the next section, we discuss related work. Then, in~\Cref{sec:Problem Definition}, we formally define the two problems that we solve. In~\Cref{sec:Solution Architecture}, we introduce solutions to both estimating the packet reception model, and indoor localization using packet counts. In~\Cref{sec:Experiment Design}, we discuss testbed set-up, collect empirical data and conduct localization experiments. We present our results in~\Cref{sec:Results} and conclude in~\Cref{sec:conclusion}.
\section{Related Work}
\label{sec:Related Work}

Now we discuss prior work related to wireless propagation models and indoor localization. Propagation models deal with loss in energy of radio waves between sender and receiver. Localization models track mobile nodes in an environment using seed nodes with known locations.

There is prior work on modeling the loss in energy during transmission for wireless signals like Wi-Fi, Bluetooth. The Received Signal Strength (RSS) i.e energy of received signal varies due to factors like distance, obstruction, walls, multi-path fading in indoor environment. \citet{zanella2016best} provides a detailed analysis of all these factors. Deterministic models like the Friis propagation model \cite{friis1946note}, Log Distance Path Loss \cite{erceg1999empirically}
give a fixed RSS value based on distance. Stochastic models like Jakes model \cite{zheng2003simulation}, two-parameter Nakagami distribution
\cite{nakagami1960m}
capture the uncertainty in received RSS values.



A wide range of techniques exist to localize a node within an indoor environment. All these techniques involve installing seed nodes in the environment with known prior location and then localizing other nodes relative to these seed nodes. These systems vary---the measuring capability of nodes, the nature of environment (i.e indoor/outdoor), and the mobility of nodes. Range-Based techniques involve the use of specialized and expensive hardware to measure some quantity which is then translated back to distance. GPS uses Time of Arrival (TOA) technique.~\citet{bahl2000radar} proposed the use of Time Difference of Arrival (TDOA) technique. Received Signal Strength based ranging techniques like SpotOn \cite{hightower2000spoton} are cheaper but inaccurate. RSS values in indoor environments becomes uncertain due to random factors like multi-path loss, fading and shadowing effects\cite{heurtefeux2012rssi}. WiFi RSS fingerprinting based methods try to mitigate this problem of inaccuracy, but require expensive human labor.~\citet{he2016wi} gives a detailed survey of all such methods.

Range free techniques do not use special hardware, but rather make assumptions on certain properties of node movement and signal propagation. Monte Carlo Localization (MCL) \cite{hu2004localization} , Mobile and Static sensor network Localization (MSL) \cite{rudafshani2007localization}, Weighted MCL \cite{zhang2010accurate} use previous location estimate and current observations to find present location of moving nodes. They assume that a heard beacon must be within a threshold distance to the current measurement location.

Our framework does not calculate RSS loss for each packet, or make any assumptions about beacon distance. Instead, we model the probability of receiving a packet, and use this probability for localization. Next we will formally define our problem and then discuss the entire solution architecture.

\section{Problem Definition}
\label{sec:Problem Definition}

Our broad goal in this work is two fold---finding a model of packet reception rate in a Bluetooth Low Energy (BLE) Internet of Things (IoT) retail store like environment and then use the model to localize individuals.

We assume that we are in a rectangular $W \times L$ space comprising stacks.
We have $k$ BLE beacons in fixed, known positions in the space. All beacons transmit at the same frequency $f$ and at the same power $r$. Further, we assume that at any location $(x,y)$, the probability of receiving packets from any beacon is binomially distributed with parameter $p$. In other words, the probability of receiving $m$ packets when we send $N$ packets is: $m \sim \operatorname{B} \left({N, p}\right)$.

\subsection{Packet Reception Rate}
\label{sub:Packet Reception Rate}
We aim to discover how $p$, the probability that we would hear a packet from a beacon varies as a function $g$ of distance ($d$), frequency ($f$) and Power ($r$). That is $p = g(d, f, r)$. Additionally g will vary based on number of intermediate stacks between beacon and packet reception location.


One can consider our probabilistic packet counting model to be a hybrid of the RSSI model and the models used in range free localization. Energy loss models
\cite{erceg1999empirically} \cite{friis1946note} are attractive in that they model the signal attenuation in the physical world. Prior work \cite{heurtefeux2012rssi} also shows that packet RSSI is highly unpredictable in indoor environment and varies with the environment layout. Existing range free models assume a spherical zone of hearing for the packets \cite{torrent2006effects} assuming that if we hear a beacon, it must be in this zone. In contrast, we make no assumptions about distance when we hear a beacon.

\subsection{Localization}
\label{sub:Localization}

Now, we list our assumptions for the localization problem. Assume that we have an individual moving in our hypothetical retail store, possessing a device that listens to the BLE beacons. This may be a smartphone, and the retail store application running on the smartphone is logging the BLE packets and then sending them to the cloud for analysis. Assume further than we would like to track the individual every $\delta$ sec. Finally, we assume stable store layout---beacon and stack locations don't change while the individual is moving.

Without loss of generality, assume that the smartphone application listens to the packets creates the following log
$$L = \{ (b_1,t_1), (b_2,t_2), \dots,  (b_N,t_N) \}$$
where $b_i$ refers to the BLE beacon \texttt{id} heard at time $t_i$. The goal is to determine a list of locations $XY_\delta=\{x_i, y_i; \delta\}$, at a store determined time resolution $\delta$ such that we know the location every $\delta$ sec.

Having presented the problems for determining packet reception and localization, we now discuss potential solutions.

\section{Solution Architecture}
\label{sec:Solution Architecture}

In this section we first show how to determine the probability of receiving a packet as a function of distance, frequency and power. Then, we present a solution to the problem of tracking individuals through the retail location using the packet reception model. Common to both approaches is a Bayesian formulation of the problem.

\subsection{Estimating the packet reception model}
\label{sub:Estimating the packet reception model}

First, we solve the problem of determining the free space packet reception model---the case when stacks are present will follow in a straightforward manner.

To determine the packet reception model,  we assume that we know the ground truth location of any spot where we listen to the beacons. Since we know the ground truth locations of all the $k$ beacons, we can calculate distance from the spot to each of the beacons heard at the spot. Assume that there exist $N$ such spots. Thus at any location $l_i, \, i \in \{1, \dots, N \}$, we have a list $D_i$ containing the number of packets of every beacon heard at $l_i$.  That is, $D_i = \{(b_j, c_j), \, j \in 1, \dots, k \}$, where $c_j$ is the count of beacon $b_j$.

We make a simplifying assumption about $g(d, f, r)$.  We assume $g$ to be an exponential function of the variable and that the log of the probability $\log p$ is quadratic in the variables. More formally:
\begin{equation}
    \log p = b_0 + \sum_i b_i x_i + \sum_{i, j} b_{i,} x_i  x_j, \quad i, j \in \{1, 2, 3\} \label{eq:linear model}
\end{equation}
where, $x_i$ refer to the variables of $d, f, r$.

Since power ($f$) and power ($r$) are constant for a specific configuration, ~\Cref{eq:linear model} reduces to a quadratic equation in distance ($d$). That is,
\begin{equation}
    \log p = b_0 + b_1 d + b_2 d^2 \label{eq:simple}
\end{equation}
The more general formulation of ~\Cref{eq:linear model} essentially states that the coefficients $b_0, b_1, b_2$ of ~\Cref{eq:simple} regress in frequency ($f$) and power ($r$). Thus the more general form allows us estimate the packet reception model for a variety of beacon power and beacon frequency configurations.

We can use Maximum Likelihood (ML) estimation via least squares to estimate the coefficients $b_i$.  We can assume that at any one of the $N$ locations, the probability $p_i$ of receiving the $i$-th beacon is:
\begin{equation}
    \bar{p_i} = \frac{c_i}{f \ast \delta},
\end{equation}
$\bar{p_i}$
where, $c_i$ is the number of packets received, $f$ is the number of packets sent per second and $\delta$ is the time window of observation. Then we can estimate $b_i$ from ~\Cref{eq:linear model} through least squares regression. The major challenge is that for low frequencies (e.g. $f = 1Hz$) or low power (e.g. $-20db$) we may not receive enough packets for a stable ML estimate of the coefficients.

A Bayesian formulation allows us to quantify the uncertainty in the coefficient estimates; when the number of packets received is large, the ML estimates and the Bayesian estimates of the coefficients will agree.

Let $\theta \equiv \{ b_i\}$ be the set of coefficients that we plan to estimate. Then the goal is estimate $P(\theta \mid D)$, where $P(\theta \mid D) \propto P(D \mid \theta) P(\theta)$. $D$ refers to the observed data---the number of packets heard for every beacon, at every location.

\begin{figure}
    \centering
    \includegraphics[width=0.4\textwidth]{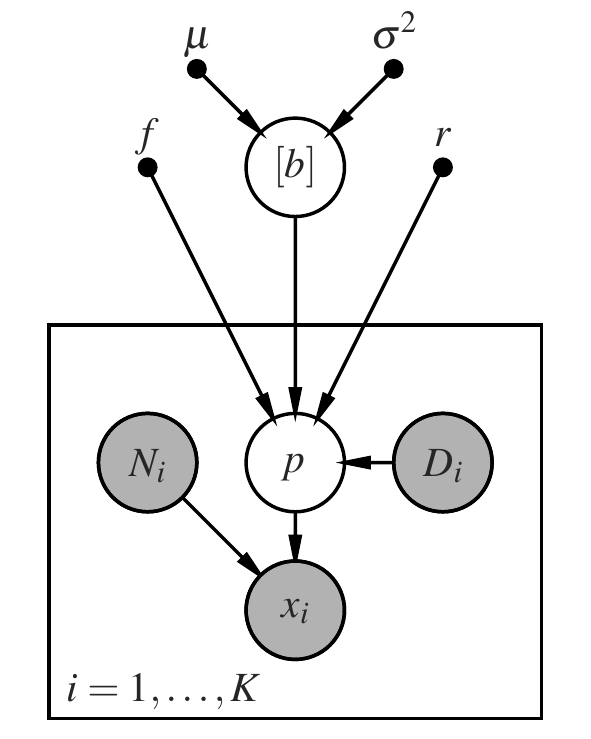}
    \caption{The figure represents the packets received at a location from each of the $K$ beacons through a generative model. The probability $p$ is a function of frequency ($f$), power ($r$), and the coefficients $b_i$. The $b_i$ are drawn from a Normal distribution $\mathcal{N}(\mu, \sigma)$. The shaded circles refer to observed variables, while the light circles refer to hidden variables, and the solid dots, parameters for $p$ and hyper-parameters for $b_i$. The plate repeats $K$ times implying that the generative prcess occurs for each of the $K$ beacons.}
    \label{fig:plate}
\end{figure}

To set up a Bayesian formulation, let us view packet reception through the lens of a generative process. Assume that we are at a particular spot $A$, listening to the $i$-th beacon. Then the number of packets received $c_i$ is drawn from a binomial distribution:
\begin{align}
    c_i &= B \left (N, p_i\right)\\
    p_i &= g(f, r, d_{i,A})
\end{align}

where the probability $p_i$ of receiving a packet from the $i$-th beacon is a function of frequency, power and the distance between the spot $A$ and the location $l_i$ of the $i$-th beacon. To formulate the priors $P(\theta)$, we assume that the prior of each of the coefficients $b_i$ is drawn from independent and identically distributed Normal distribution.  That is,
\begin{equation}
    b_i \sim \mathcal{N}(\mu, \sigma),
\end{equation}
where we set $\mu=0$ and $\sigma=10$ so that the priors are conservative, allowing for a large range of values.
The Bayesian formulation is compactly summarized in ~\Cref{fig:plate}.
We compute the posterior $P(\theta \mid D)$ using a standard Markov Chain Monte Carlo technique.

We can use the same formulation for the free space case and the case when there are stacks. At each location, we filter the packets based on the beacon \texttt{id} allowing us to separately analyze the different cases since we know the ground truth location of the beacons and we know their distances to the location where we are making the measurement.

\subsection{Estimating location sequence}
\label{sub:Estimating location sequence}

We use Bayesian formulation for determining the location of a person in a store. We plan to use the layout of the space to impose constraints on the solution.


Let us begin with what is observable. As before, at any location, the observations include the packet counts from each beacon. We do know the ground truth locations of each beacon, the frequency ($f$) of transmission and the power ($r$).
Now due to the results of ~\Cref{sub:Estimating the packet reception model}, we know the parameters of the packet reception model.

For any location, we need to estimate  hidden parameters. First, since we don't know the location, \textit{we don't know the location of any of the beacons relative to the current position}. We do not know, when we receive packets from the $i$-th beacon, if the $i$-th beacon is in the same aisle, or one or more aisles away. Thus the number of aisles between the current location and any beacon is a latent parameter for that beacon. The speed $s$ at which a person moves through the store is a latent variable. We can assume an upper bound for the speed.

Now, we describe the movement model. Let us assume as before that we wish to estimate the true $(x,y)$ values at $N$ locations, where the $N$ depends on  the temporal resolution at which the retail store wishes to track its customers. The basic movement model assumes the following priors:
\begin{align*}
    s_i & \sim U(0, S_{max}), \, i \in \{1, \dots, N-1 \} \\
    x_0 &\sim U(0, W),\\
    y_0 & \sim U(0, L),\\
    x_i\mid x_{i-1} &\sim \mathcal{N}(0, s_{i-1}*\delta), \, i \in \{1, \dots, N-1 \} \\
    y_i \mid y_{i-1} &\sim \mathcal{N}(0, s_{i-1}*\delta), i \in \{1, \dots, N-1 \} .
\end{align*}

\vspace{-3.5pt}
Where, $s_i$ refer to the speed between locations, and uniform prior until some speed $S_{max}$, $(x_0, y_0)$ are the initial $(x,y)$ locations of the person, and since we know little about them, we assume that they are uniformly distributed over the space. We assume that an intermediate location $(x_i, y_i)$ is Normally distributed around $(x_{i-1}, y_{i-1})$ with a standard deviation equal to $s_{i-1} * \delta$, where $s_{i-1}$ is the speed with which the person left the previous location $(x_i, y_i)$ and where $\delta$ is the time window of observation.

To estimate the location of the beacon relative to the measurement location, we make use of the layout of the space. We arrange our beacons in regularly spaced intervals on stacks. The beacons on the two sides of an aisle form a group. In ~\Cref{image9}, beacon numbers [1-12], [13-36] and [37-60] form three groups. All beacons in the same group must be an identical number of stacks away from the current location. Thus all beacons in the same group will use the same packet reception model. In our layout, the packet reception model used for a beacon group will depend on the $y$ coordinate of the person. We can model the decision to switch as follows:
\begin{align}
    \tau_i &\sim U(0, L), \, i \in \{1, 2\} \nonumber\\
    A_i &= \begin{cases}
    0 \quad y_i < \tau_1,\\
    1 \quad \tau_1 \leq y_i \leq \tau_2, \\
    2 \quad y > \tau_2, \end{cases} \label{eq:cases}\\
    S_{i,k} &= M(A_i, b_k). \nonumber
\end{align}

Where, $\tau_i$ are two latent variables with a uniform prior along the $y$ direction; ~\Cref{eq:cases} helps us determine the estimate of the current aisle $A_i$ and $M$ is a deterministic mapping of the relative number of the stacks $S_{i,k}$ between the current location $i$ and beacon $k$. We can do this mapping because we know the store layout. The variable $S_{i,k}$ helps us determine the appropriate packet reception model.

We estimate parameters $\theta \equiv \{ \{x_i, y_i \}, \{s_i\}, \tau_i \}$.  The data collected $D$ over all locations include the packet counts $\{ c_k\}$ of each beacon $b_k$ within each time window. Notice that since we estimate $S_{i,k}$ the number of stacks between beacon $k$ and current location $i$, we use the following relation:
\begin{align*}
    c_{i,k} &= B(M, p_{i,k}), \, M= f * \delta,\\
    p_{i,k} &= g \left (f, r, d_{i,k; \, S_{i,k}} \right),\\
    d_{i,k} &= \sqrt{(x_i - b_{k,x})^2 + (y_i - b_{k,y})^2}.
\end{align*}

Where, the packet counts $c_{i,k}$ of the $k$-th beacon at location $i$ is Binomially distributed with parameter $p_{i,k}$. We obtain the parameter $p_{i,k}$ using the correct packet reception model, by using the estimate of the number of stacks $S_{i,k}$ between location $i$ and location of beacon $k$. The distance between the location $i$ and location of beacon $k$ denoted as $(b_{k,x}, b_{k,y})$ is the standard Euclidean distance.

Our goal is to estimate $P(\theta \mid D) \propto P(D \mid \theta) P(\theta)$. We use a standard MCMC framework to estimate $P(\theta \mid D)$.

What if the store geometry was not so simple to use the two latent random variables $\tau_i$?. We can formulate the number of stacks between the beacon and the location in a more general way using a Dirichlet ditribution as a prior:

\vspace{-2pt}

\begin{align*}
q_{i,k} &\sim \mathrm{Dir}(\alpha)\\
S_{i,k} &\sim \mathrm{Cat}(q_{i,k})
\end{align*}

\vspace{-4pt}
Where we use a symmetric Dirichlet distribution with parameter $\alpha=1$; We draw a three dimensional distribution $q_{i,k}$ from the Dirichlet, for each location $i$ and for each beacon $k$ corresponding to the probabilities that there is either no stack, or one stack or  two stacks respectively, between beacon $k$ and location $i$. We would use probabilities $q_{i,k}$ to then draw from a categorical distribution. We did not use this formulation, since in our case we could exploit geometric constraints.

In this section we presented a solution to estimating the packet reception model and then showed how to use that model in locating an individual as she walks in a retail environment. A Bayesian formulation is central to solving both problems. In the next section, we discuss how we gathered empirical data to develop our packet reception model model and how we use the developed model to locate the individual.

\section{Experiment Design}
\label{sec:Experiment Design}

In this section we will describe the three steps of carrying out the real world experiments---setting up the devices (\Cref{sub:Device Set-Up}), the experimental testbed (\Cref{sub:Environment Set-Up}) and data collection (\Cref{sub:Data Collection}).
\subsection{Device Set-Up}
\label{sub:Device Set-Up}

First we discuss three device types used in our testbed --- Bluvision iBeeks, BluFi, TI packet sniffer.

iBeeks send out bluetooth low energy (BLE) packets into the environment and act as seed nodes of location. We choose these particular beacons because of their battery capacity, transmission power range and high advertising frequency. Their batteries last for a long time ranging from three to nine years. They support a wide range of broadcasting power from -40 dBm to +5 dBm. -40 dBm translates to 3 meter line of sight range, while +5 dBm gives us a range as large as 150 meter. We test the impact of range of sight on localization accuracy in our experiments. The beacons advertise packets as fast as one per 100 milliseconds. iBeeks are installed on particular locations in the environment and they remain stationary throughout the experiment. As their locations are known to us, they act like seed nodes based on which the location of other nodes are estimated.

BluFi enables mass re-configuration of iBeeks. To test the effects of frequency and power on packet reception rate, we need to re-configure the beacons at regular intervals. Bluzone app allows us to talk with single iBeek at a time. BluFi pushes new configurations to thousands of beacons with one single command from the Bluzone cloud.
Thus this device proves to be essential in large scale BLE beacon deployments.

Texas Instrument Packet Sniffer scans BLE packets sent out by iBeeks and also act as the node for which we want to estimate the location. iBeeks broadcast on three different channels and the packet properties vary a lot based on the channel.
The sniffer is a CC2540 dongle developed by Texas Instruments that can capture BLE packets on one advertising channel. The packets captured can be shown in real time by the Smart RF Packet Sniffer Software.
 The sniffer connected to a Windows laptop is kept at fixed locations during the training phase to collect the beacon packet trace. We walk around with the sniffer during the test phase to collect movement traces.

\subsection{Environment Set-Up}
\label{sub:Environment Set-Up}

Now we will report on two environments that constitute our testbed---Undergraduate Library (UGL) and Grainger Engineering Library at the University of Illinois at Urbana-Champaign. Both environments are subareas of a library floor. They have book shelves segregating the floor into aisles and corridors. We chose to experiment in library spaces since we didn't have ready access to retail locations; we hope to perform future experiments in actual retail stores. The floor plan is like retail stores where we have stack of items. The two environments differ: presence of walls, different kinds of obstructing materials.

We do the training phase of the experiment at the UGL. This phase involves collecting of packet trace data at different locations. We  estimate the packet reception model parameters using empirical data collected at this location. Aisles between shelves provide free space and they are 1.22 meters wide. We use two bookshelves, each 0.64 meters wide and 17 meters long. On each aisle, we place two rows of 16 beacons on the two shelves facing the aisle. The inter-beacon distance on the same row is 1 meter, while the inter-beacon distance for beacons on the same shelf, but on different aisles is 0.64 meter i.e the thickness of the book shelf. The shelves are made of wood. We collect the packet traces in the aisles.

The testing phase takes place at the Grainger Library. This phase involves using the packet reception model to localize a moving person in the space. The Grainger environment differs from the training phase location in three  aspects. First, there are steel bookshelves as opposed to wooden shelves in UGL. Second, there is more open space on either side of boundary shelves as opposed to a more closed feature with walls on either side in UGL. We expect the effects of multi-path fading to be different. Third, this particular region has high foot traffic people in contrast to the training location where foot traffic was low. This will help us study the impact of dynamic human presence on localization.

The testing location differs in number of stacks, length and width of the aisle. Each stack is 11 meters long and 0.5 meters wide. The environment comprises three such stacks. Aisles are 0.7 meters wide. We place two rows of 12 beacons on each stack. The inter-beacon distance on the same row is 0.91 meter, while the inter-beacon distance for two devices kept opposite each other on the same shelf, but facing two different aisles is 0.43 meter.


\subsection{Data Collection}
\label{sub:Data Collection}

We collect two types of data at different power and frequency---beacon packet trace required for training the packet reception model and movement trace to test the utility of the packet reception model in localization.

We collect beacon packet traces during the training phase while standing at fixed spots in the layout. Since the distance calculations have to be exact, we do not introduce mobility in this step.
The broad steps for this phase are the following.
\begin{enumerate}
\item Placing the beacons on the shelves at regular intervals.
\item Using BluFi to re-configure the beacons to desired parameter settings (power, advertising  frequency).
\item Collecting the packet trace for current beacon configuration at three fixed locations per aisle. Two locations chosen near the two ends of each aisle and one in the middle.
\item Repeating Steps 2 and 3 until all the desired parameter settings are covered.
\end{enumerate}

\begin{figure}[htbp]
    \centering
    \includegraphics[viewport=0.8in 1.3in 11.1in 11.3in, width=\columnwidth,clip=true]{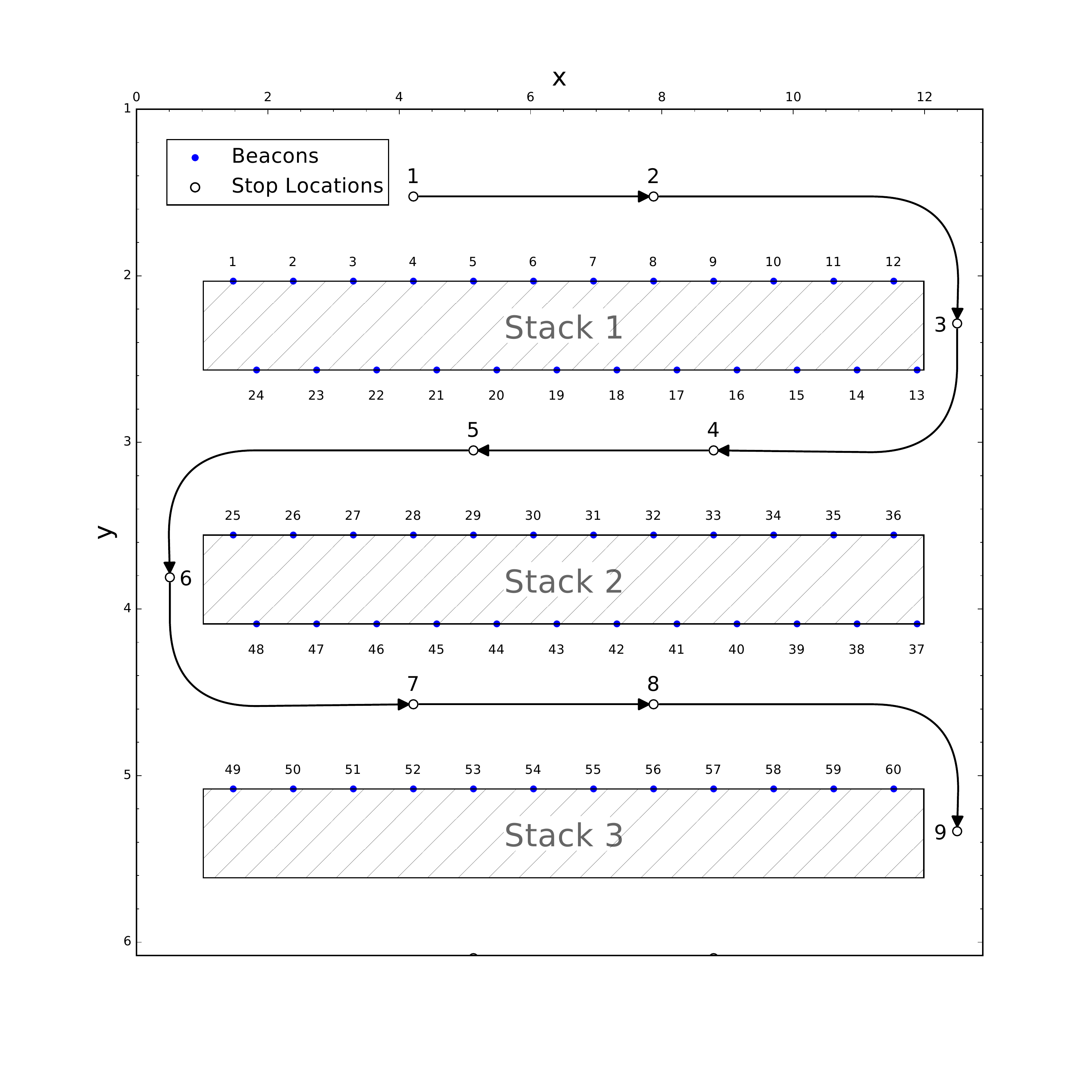}
    \caption{The test environment layout with three stacks. We show the movement Sequence shown by the curve starting at stop location 1 and ending at 9. The stop locations act as destination in our modified random waypoint movement model.}
    \label{image9}
\end{figure}

We collect movement trace in the testing phase by carrying out a modified random waypoint mobility model. This trace data contains two parts---packet trace heard during movement and actual ground truth locations. Obtaining actual locations while moving becomes is a challenge. We address the challenge by carrying out a random waypoint like movement model in real world with one modification: we fix in advance all the destinations while introducing movement randomness. Stop locations marked in ~\Cref{image9} act as destinations. We start moving from one end aisle and finish in the other. ~\Cref{image9} shows the exact movement sequence at the testing location starting from stop location 1 and ending at 9. Like a waypoint model, the speed of movement remains random since an actual person is doing the movement. The pause time after reaching each destination is also randomly chosen between 8 seconds and 10 seconds. We collect the movement trace for all beacon parameter settings. After one round of movement we use BluFi to re-configure all the beacons.


\section{Results}
\label{sec:Results}

In this section we estimate the packet reception model parameters in~\Cref{sub:Inferring the Noise Model} and the localization using the packet reception model in~\Cref{sub:Localization Accuracy}.
\subsection{Inferring the Noise Model}
\label{sub:Inferring the Noise Model}

The variables affecting the packet reception rate are distance, frequency and beacon power. We measure distance, represented as ($d$), in meters and frequency, shown as ($f$), in Hertz (Hz). 1Hz advertising frequency represents a time interval of 1 sec between each packet. We represent beacon power in dBm. Since dBm is a relative figure, we use -12dbm as a reference to compute the parameters in our model. Our reference power of -12dBm translates to a 10-12m beacon hearing range.


\begin{figure}[!h]
    \centering
    \includegraphics[viewport=0.98in 1.2in 13.1in 14.1in, width=\columnwidth,clip=true]{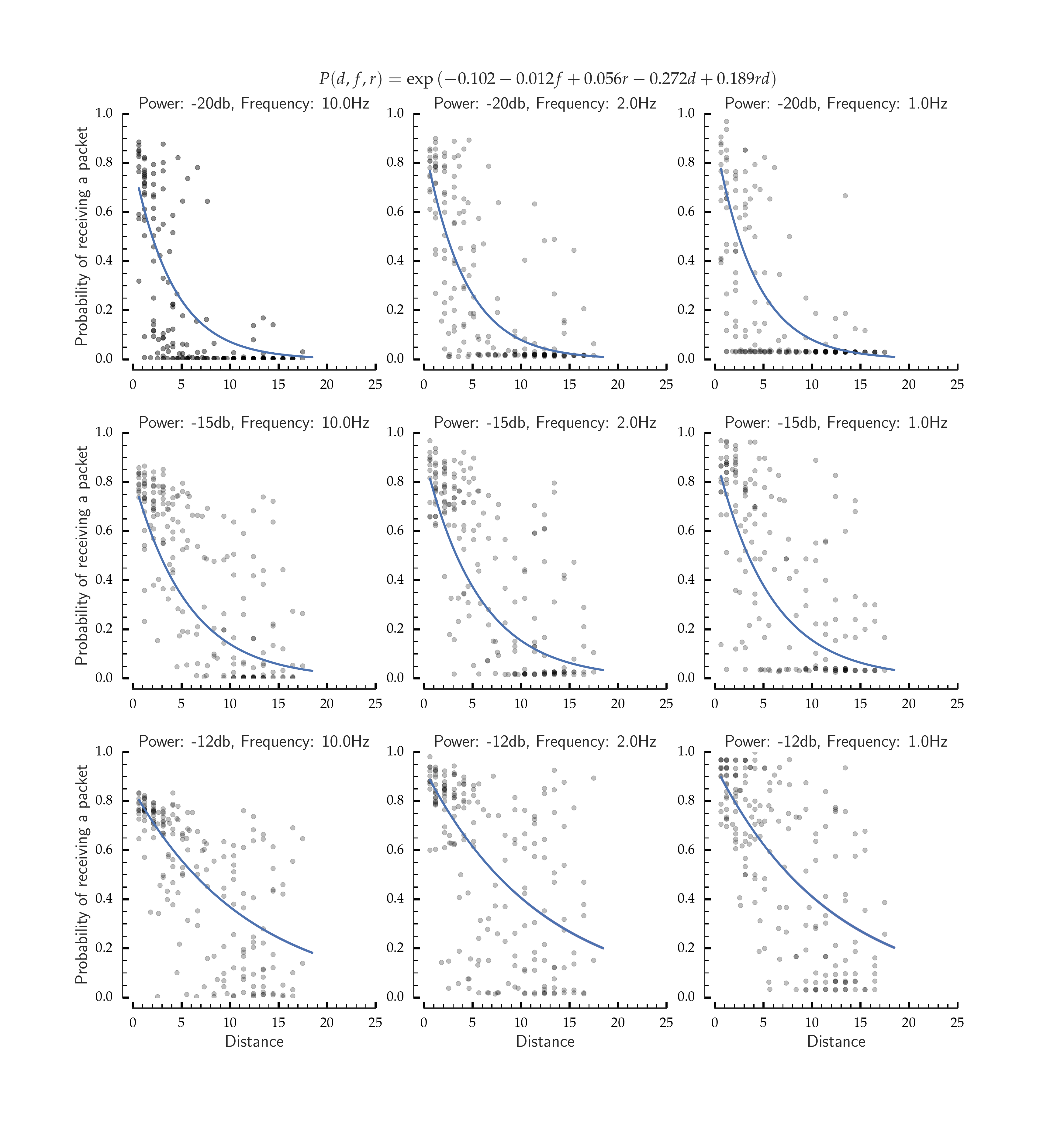}
    \caption{The figure show the generalized linear model for the free space case, fit to changing values of frequency and power. Notice that packet reception increases with decreasing frequency and with power.} \label{fig:glm}
\end{figure}

We collect data at three values each for device parameters of frequency and power. We use frequency values of 1Hz, 2Hz and 10Hz. High frequency of 10Hz helps us to check the effect of high packet emission rate on the noise or confusion in the medium. Such noise in turn can lead to lower reception rate. We set beacon powers at values -20db, -15db, -12db. -12db gives us a large range of 10-11 meters which  almost covers our entire experiment layout. -20db covers a much smaller range of 3-4 meters. We carried out experiments and collected data at all nine possible combinations of these two parameters.

\begin{figure}[t]
\includegraphics[viewport=0.5in 0.2in 9.8in 17.8in, width=\columnwidth,clip=true]{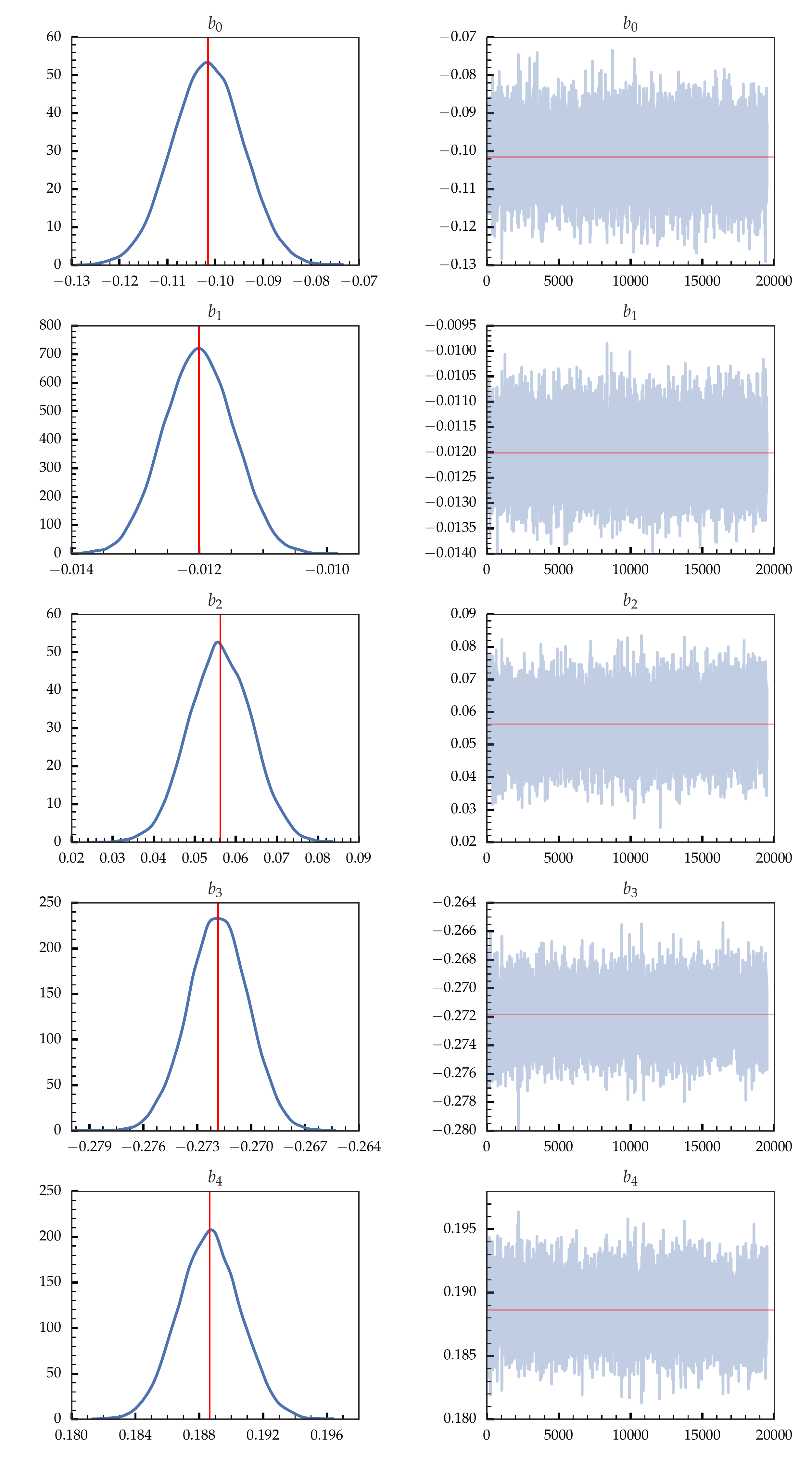}
\caption{ Posterior Distribution of General Model parameters for the free space case, with the vertical lines showing the mean. The posterior distributions are converging}
\label{fig:glmpost}
\end{figure}




We estimate the posterior $P(\theta \mid D)$ using PyMC3 a standard MCMC package~\cite{Salvatier2016}. We estimate $\theta$ (parameter values $b_i$ of~\Cref{eq:linear model}) by using the entire dataset that includes all nine combinations of frequency and power.
~\Cref{fig:glmpost} shows the distributions. The plot shows that the distribution for all the coefficients have converged.
Taking the mean estimates of the posterior distributions of each parameter, the model for  $\log p_0$ the $\log$ of the free space packet reception probability:
$$ \log p_0 = -0.101 - 0.012f + 0.056r -0.272d + 0.189rd$$
The mean of the coefficients of $d^2, r^2, f^2, f\cdot d$ are close to zero and ignored.

~\Cref{fig:glm} shows the free space model fit to the raw data for all nine combinations of power and frequency. The dots show the raw packet counts received at varying distance while the curves represent the Bayesian fit. We can see from the figure that increasing power increases the packet reception probability and that \textit{decreasing} frequency \textit{increases} packet reception due to decreasing packet interference.



\begin{figure}[t]
    \centering
\includegraphics[viewport=1.1in 0.02in 13.4in 4.91in, width=\columnwidth,clip=true]{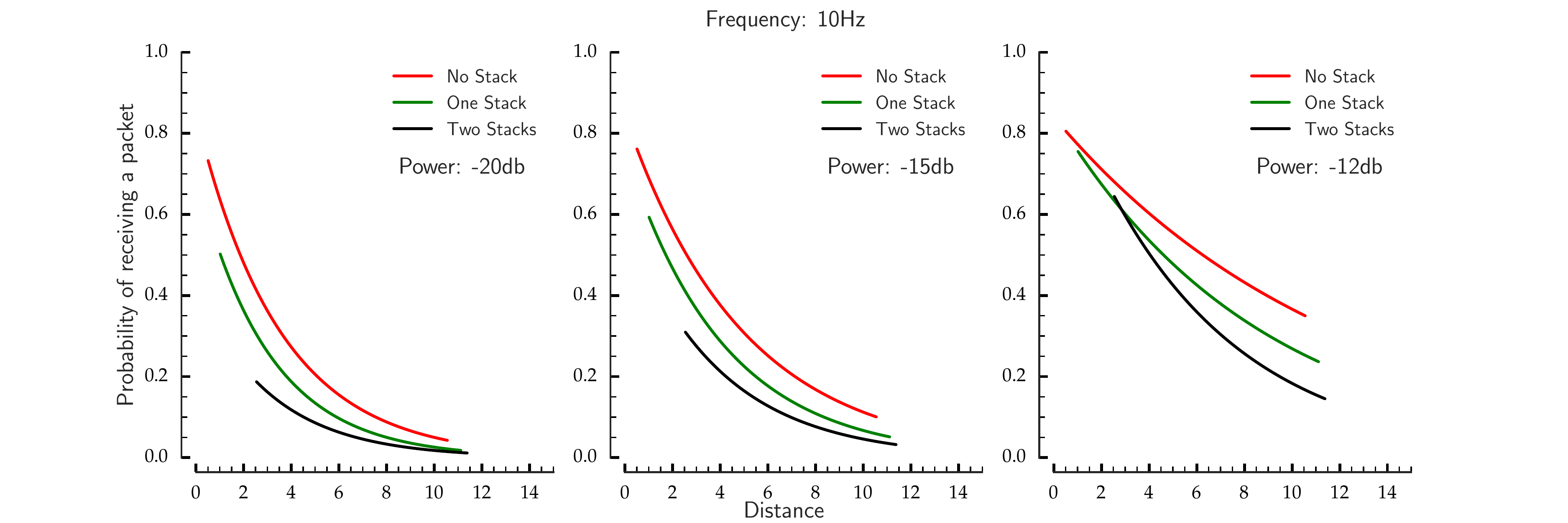}
\caption{ Generalized model fit including stacks ($f=10Hz$). Reception decays due to the presence of stacks. Increasing power from left to right has most effect for two stacks followed by one stack and the least effect for no stacks (free space). An increase in power helps to overcome dampening due to stacks.}
\label{fig:glmall}
\end{figure}


Now we infer a stack model where the process of estimation remains the same, but we filter the data points such that there is obstruction present between the device and receiver.
Figure \ref{fig:glmall} shows a comparison of the different stack models for a fixed frequency of 10Hz. One stack and two stack model obtained on estimation is as follow.
\begin{align*}
 \log p_1 &=  -0.236 - 0.026f + 0.303r -0.292d + 0.018rd\\
 \log p_2 &=  -0.305 - 0.033f + 0.604r -0.302d + 0.017rd
\end{align*}
where $p_i, \, i \in \{1,2\} $ is the probability for the case of one stack and two stacks respectively, and where, $f, r, d$ represent frequency, power and distance respectively.

With the increase in the number of stacks, the constant factor in packet reception becomes lower (i.e parameter $b_0$ becomes more negative). This means in general, we have a lower chance of getting a packet. Similarly, the decay rate due to frequency $b_1$ and distance $b_3$ increase as well. The most significant change occurs in the impact of beacon power on packet reception. The coefficient of r, $b_2$ jumps from 0.056 in the no stack case to 0.303 in one stack case and 0.604 in two stack case. This is also evident in~\Cref{fig:glmall} where the gradual increase of power from left to right has more impact on two stack and one stack cases as compared to the no stack case. We can justify this result by the fact that the stacks dampen the power of the transmitted packets and larger power helps in crossing this barrier leading to higher reception. Beacon power plays more significant role in reception across stacks. Due to increased role of beacon power in overcoming the stacks, it has less impact on compensating for distance which is evident by decreasing value of $b_4$.

Thus, packet reception varies based on distance, frequency, power and presence of obstructions. It decreases with increase in distance, frequency or number of obstructions. Power plays a vital role in compensating for the effects of both distance and obstructions. It helps in increasing reception across obstructions and to a larger distance in free space.

\subsection{Localization Accuracy}
\label{sub:Localization Accuracy}

In this section we present the accuracy using our packet reception probability model along with MCMC localization. We term our localization framework as Packet Count-Monte-Carlo Localization or PC-MCL in short. We compare against a standard range free localization algorithm, MCL \cite{hu2004localization} to see the effects of the packet reception model on its performance. While more recent work \cite{rudafshani2007localization}, \cite{zhang2010accurate} improve upon the standard MCL accuracy, all assume a hard threshold model for hearing the beacons (i.e. if they hear a beacon, then  it must be nearer some threshold distance $d_0$).

\begin{table}[!h]
  \begin{tabular}{cccc}
    \toprule
    Power & Frequency & PC-MCL error ($m$) & MCL error ($m$)\\
    \midrule
    -20dB & 2 Hz & 1.99 $(\downarrow40.2\%)$ & 3.33\\
    -20dB & 1 Hz & 1.83 $(\downarrow45.4\%)$ & 3.35\\
    -15dB & 10 Hz & 1.11 $(\downarrow65.3\%)$ & 3.20\\
    -15dB & 2 Hz & 1.48 $(\downarrow53.3\%)$ & 3.17\\
    -15dB & 1 Hz & 1.39 $(\downarrow57.9\%)$ & 3.30\\
    -12dB & 2 Hz & 1.56 $(\downarrow54.1\%)$ & 3.40\\
    -12dB & 1 Hz & 1.49 $(\downarrow54.8\%)$ & 3.30\\
  \bottomrule
\end{tabular}
\caption{Average estimation error in meters for the proposed Packet Count-MCL against the standard MCL. The PC-MCL error varies in range $1-2m$, while the standard MCL error is always over $ 3m$. Error is least for $-15dB$ power. In a sense, $-15db$ is ``just right'': $-20dB$ has low beacon coverage of physical space and $-12dB$ increases confusion with high coverage.}
\label{errTable}
\end{table}


We estimate location in discrete time intervals of size $\delta$ and then calculate localization error over each interval. We segregate the movement trace of the person into time windows each of duration $\delta$ seconds. In our case, we choose $\delta=10$ sec. Localization error in each interval is the euclidean distance between predicted and ground truth location.

~\Cref{errTable} shows the average error for different device settings. Packet count based MCL gives higher accuracy compared to baseline MCL. Our system can localize within a range of $1-2m$ while baseline MCL always has an error over $3m$. Note that the errors are lowest for a device power of $-15dB$. This is because at $-20dB$ power beacons have a low coverage and individual moving in the space may not receive sufficient number of packets to get localized with low error. In contrast, $-12dB$ gives high coverage and we hear all the beacons with increased reception rate throughout our layout. This makes it slightly harder to distinguish through which aisle the person is moving.

\begin{figure}[!h]
    \centering
\includegraphics[viewport=1.0in 0.20in 11in 5.7in, width=\columnwidth,clip=true]{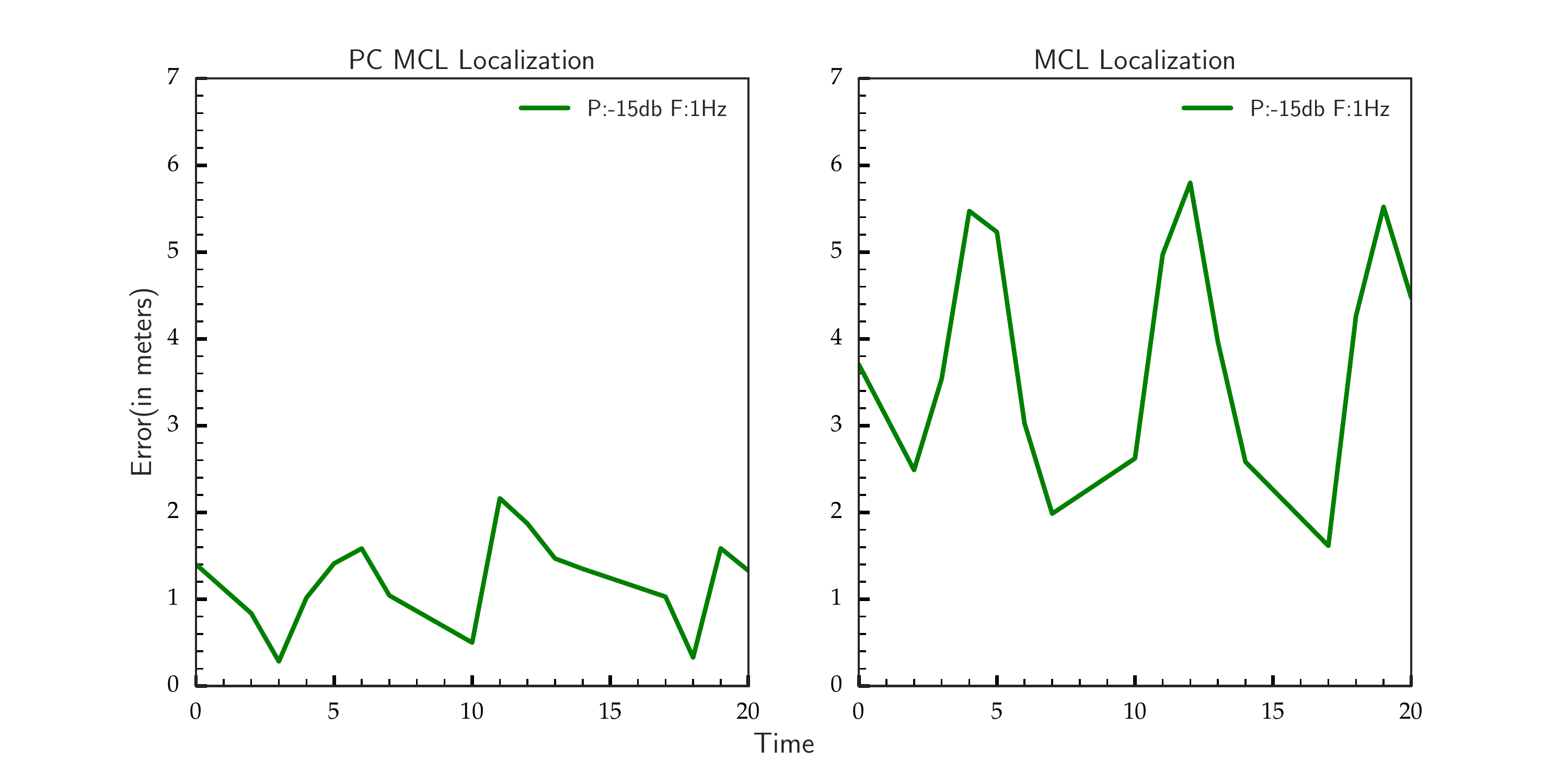}
\caption{Localization errors over time for both PC-MCL and standard MCL. Standard MCL average error is around 0.4*radio range,  consistent with \cite{hu2004localization}. Radio range is 10 meter for -15db. Errors increase during time intervals 4-5-6 and 10-11-12 because the person is transitioning between aisles.}
\label{image7}
\end{figure}

The average localization error within an aisle, and when the person transitions between the aisles using the corridor are different.  ~\Cref{image7} shows the time series variation of error of our proposed PC-MCL and the baseline MCL algorithms for a device setting of $-15dB$ and $1Hz$. We see that the errors increase during the time intervals $4-5-6$ and $10-11-12$ for both the algorithms. This is because during the transition we don't have the right packet reception model to be used. Thus the average error of both algorithms increases due to errors during the transition. Indeed, the average error within an aisle drops to as low as $0.4m$ with our PC-MCL algorithm. Thus, if we can eliminate the high errors during transition, our algorithm can achieve high localization accuracy in the range of  $0.4-0.5m$. One way to achieve this to learn a packet counting model for the corridor where transitions occur, in addition to the packet count model for the aisles.

\section{conclusion}
\label{sec:conclusion}

In this paper, we developed a probabilistic model for BLE packet reception in an indoor environment with stacks and used this model to localize moving individuals in an indoor environment. We observed that the packet counts for a beacon are binomially distributed with a parameter $p$, and then modeled $p$ as function of advertising frequency, beacon power and distance to beacon. We estimated the coefficients using a Bayesian MCMC technique. We developed a Monte-Carlo localization technique using the packet reception model exploiting environment geometry in our solution. Our proposed framework performs well: we achieve an average reduction of $53\%$ in localization error compared to a baseline Monte-Carlo localization algorithm.

We can improve our proposed framework. We noticed that while our average localization error was around $\sim 1.2m$, the errors \textit{within} an aisle were $\sim 0.4m$. The increase in the average localization error is due to poor localization during the transition. This leads us to conclude that a ``corridor'' packet model in conjunction to the proposed ``aisle'' packet model will lead to reduction of average localization error.

\bibliographystyle{ACM-Reference-Format}
\bibliography{ref}

\end{document}